\documentclass[5p]{elsarticle}
\usepackage{hyperref}

\usepackage[utf8]{inputenc}
\usepackage{amsmath}
\usepackage{caption}
\usepackage{subcaption}
\usepackage{float}
\usepackage[T1]{fontenc}
\usepackage{xcolor}
\usepackage{multirow}
\biboptions{numbers,sort&compress}
\graphicspath{{Images/}} 

\begin{document}

\begin{frontmatter}

\title{Production Rate Measurement of Tritium and Other Cosmogenic Isotopes in Germanium with CDMSlite}

\author[a]{R.~Agnese} 
\author[b]{T.~Aralis} 
\author[c]{T.~Aramaki} 
\author[d]{I.J.~Arnquist} 
\author[e]{E.~Azadbakht} 
\author[e]{W.~Baker} 
\author[f]{S.~Banik} 
\author[g]{D.~Barker} 
\author[h]{D.A.~Bauer} 
\author[i]{T.~Binder} 
\author[j]{M.A.~Bowles} 
\author[c]{P.L.~Brink} 
\author[d]{R.~Bunker} 
\author[k]{B.~Cabrera} 
\author[l]{R.~Calkins} 
\author[c]{C.~Cartaro} 
\author[m,n]{D.G.~Cerde\~no} 
\author[b]{Y.-Y.~Chang}
\author[l]{J.~Cooley} 
\author[b]{B.~Cornell}
\author[g]{P.~Cushman} 
\author[p]{T.~Doughty} 
\author[o]{E.~Fascione\corref{cor1}} 
\ead{efascione@owl.phy.queensu.ca}
\cortext[cor1]{Corresponding author}
\author[q]{E.~Figueroa-Feliciano} 
\author[p]{C.W.~Fink} 
\author[g]{M.~Fritts} 
\author[o]{G.~Gerbier} 
\author[o]{R.~Germond} 
\author[o]{M.~Ghaith} 
\author[b]{S.R.~Golwala} 
\author[e,r]{H.R.~Harris} 
\author[q]{Z.~Hong} 
\author[d]{E.W.~Hoppe} 
\author[h]{L.~Hsu} 
\author[s,t]{M.E.~Huber} 
\author[f]{V.~Iyer} 
\author[l]{D.~Jardin}
\author[e]{A.~Jastram}
\author[f]{C.~Jena} 
\author[c]{M.H.~Kelsey} 
\author[g]{A.~Kennedy} 
\author[e]{A.~Kubik} 
\author[c]{N.A.~Kurinsky} 
\author[e]{R.E.~Lawrence} 
\author[d]{B.~Loer} 
\author[m]{E.~Lopez~Asamar} 
\author[h]{P.~Lukens} 
\author[u,v]{D.~MacDonell} 
\author[e]{R.~Mahapatra} 
\author[g]{V.~Mandic} 
\author[g]{N.~Mast} 
\author[j]{E.~Miller} 
\author[e]{N.~Mirabolfathi} 
\author[f]{B.~Mohanty} 
\author[e]{J.D.~Morales~Mendoza} 
\author[g]{J.~Nelson} 
\author[d]{J.L.~Orrell}
\author[u,v]{S.M.~Oser} 
\author[u,v]{W.A.~Page} 
\author[c]{R.~Partridge} 
\author[g]{M.~Pepin}
\author[k]{F.~Ponce} 
\author[i]{S.~Poudel} 
\author[p]{M.~Pyle} 
\author[l]{H.~Qiu} 
\author[o]{W.~Rau} 
\author[w]{A.~Reisetter} 
\author[q]{R.~Ren} 
\author[a]{T.~Reynolds} 
\author[s]{A.~Roberts} 
\author[x]{A.E.~Robinson} 
\author[g]{H.E.~Rogers} 
\author[a]{T.~Saab} 
\author[p,y]{B.~Sadoulet} 
\author[i]{J.~Sander} 
\author[u,v]{A.~Scarff} 
\author[j]{R.W.~Schnee} 
\author[z]{S.~Scorza} 
\author[f]{K.~Senapati} 
\author[p]{B.~Serfass} 
\author[p]{D.~Speller} 
\author[l]{M.~Stein} 
\author[j]{J.~Street} 
\author[c]{H.A.~Tanaka} 
\author[e]{D.~Toback} 
\author[o]{R.~Underwood} 
\author[s]{A.N.~Villano} 
\author[u,v]{B.~von~Krosigk} 
\author[p]{S.L.~Watkins} 
\author[e]{J.S.~Wilson} 
\author[aa]{M.J.~Wilson} 
\author[e]{J.~Winchell} 
\author[c]{D.H.~Wright} 
\author[k]{S.~Yellin} 
\author[ab]{B.A.~Young} 
\author[o]{X.~Zhang} 
\author[e]{X.~Zhao}

\address[a]{Department of Physics, University of Florida, Gainesville, FL 32611, USA}
\address[b]{Division of Physics, Mathematics, \& Astronomy, California Institute of Technology, Pasadena, CA 91125, USA}
\address[c]{SLAC National Accelerator Laboratory/Kavli Institute for Particle Astrophysics and Cosmology, Menlo Park, CA 94025, USA}
\address[d]{Pacific Northwest National Laboratory, Richland, WA 99352, USA}
\address[e]{Department of Physics and Astronomy, and the Mitchell Institute for Fundamental Physics and Astronomy, Texas A\&M University, College Station, TX 77843, USA}
\address[f]{School of Physical Sciences, National Institute of Science Education and Research, HBNI, Jatni - 752050, India}
\address[g]{School of Physics \& Astronomy, University of Minnesota, Minneapolis, MN 55455, USA}
\address[h]{Fermi National Accelerator Laboratory, Batavia, IL 60510, USA}
\address[i]{Department of Physics, University of South Dakota, Vermillion, SD 57069, USA}
\address[j]{Department of Physics, South Dakota School of Mines and Technology, Rapid City, SD 57701, USA}
\address[k]{Department of Physics, Stanford University, Stanford, CA 94305, USA}
\address[l]{Department of Physics, Southern Methodist University, Dallas, TX 75275, USA}
\address[m]{Department of Physics, Durham University, Durham DH1 3LE, UK}
\address[n]{Instituto de F\'{\i}sica Te\'orica UAM/CSIC, Universidad Aut\'onoma de Madrid, 28049 Madrid, Spain}
\address[o]{Department of Physics, Queen's University, Kingston, ON K7L 3N6, Canada}
\address[p]{Department of Physics, University of California, Berkeley, CA 94720, USA}
\address[q]{Department of Physics \& Astronomy, Northwestern University, Evanston, IL 60208-3112, USA}
\address[r]{Department of Electrical and Computer Engineering, Texas A\&M University, College Station, TX 77843, USA}
\address[s]{Department of Physics, University of Colorado Denver, Denver, CO 80217, USA}
\address[t]{Department of Electrical Engineering, University of Colorado Denver, Denver, CO 80217, USA}
\address[u]{Department of Physics \& Astronomy, University of British Columbia, Vancouver, BC V6T 1Z1, Canada}
\address[v]{TRIUMF, Vancouver, BC V6T 2A3, Canada}
\address[w]{Department of Physics, University of Evansville, Evansville, IN 47722, USA}
\address[x]{D\'epartement de Physique, Universit\'e de Montr\'eal, Montr\'eal, Qu\'ebec H3C 3J7, Canada}
\address[y]{Lawrence Berkeley National Laboratory, Berkeley, CA 94720, USA}
\address[z]{SNOLAB, Creighton Mine \#9, 1039 Regional Road 24, Sudbury, ON P3Y 1N2, Canada}
\address[aa]{Department of Physics, University of Toronto, Toronto, ON M5S 1A7, Canada}
\address[ab]{Department of Physics, Santa Clara University, Santa Clara, CA 95053, USA}

\begin{abstract}
Future direct searches for low-mass dark matter particles with germanium detectors, such as SuperCDMS SNOLAB, are expected to be limited by backgrounds from radioactive isotopes activated by cosmogenic radiation inside the germanium. There are limited experimental data available to constrain production rates and a large spread of theoretical predictions. We examine the calculation of expected production rates, and analyze data from the second run of the CDMS low ionization threshold experiment (CDMSlite) to estimate the rates for several isotopes. We model the measured CDMSlite spectrum and fit for contributions from tritium and other isotopes. Using the knowledge of the detector history, these results are converted to cosmogenic production rates at sea level. The production rates in atoms/(kg$\cdot$day) are 74\,$\pm$\,9 for $^3$H, 1.5\,$\pm$\,0.7 for $^{55}$Fe, 17\,$\pm$\,5 for $^{65}$Zn, and 30\,$\pm$\,18 for $^{68}$Ge.
\end{abstract}


\end{frontmatter}


\section{Introduction}

Astrophysical observations indicate that dark matter constitutes a majority of the matter in the Universe \cite{:ParticleReview,:planck}. Weakly interacting massive particles (WIMPs) are a well-motivated class of candidates that could explain these observations \cite{:steigman,:lee} and may be directly detectable with a sufficiently sensitive Earth-based detector \cite{:Goodman}. Traditionally, direct searches have focused on WIMPs with masses in the range of $\sim$10\,GeV/c$^2$ to several TeV/c$^2$. Although searches in this mass range are ongoing, the lack of evidence for such particles~\cite{:Akerib,:Aprile,:pandaii}, or for supersymmetry at the Large Hadron Collider~\cite{:ATLAS,:CMS}, motivates exploration of lower-mass alternatives~\cite{:LowMass1,:LowMass2,:LowMass3,:LowMass4,:LowMass5}.

The kinematics of low-mass dark matter interactions with atomic nuclei lead to low energy nuclear recoils (NRs). The performance of discrimination techniques typically used to distinguish electron-recoil (ER) background from NRs generally degrades with decreasing recoil energy \cite{:HT,:Aprile,:pico,:DEAP,:cresst}. The ER background is therefore likely to become the primary limiting factor for the experimental reach of low-mass dark matter searches \cite{:sensitivity}. A particularly important source of ERs is radioactivity produced through cosmogenic activation of the detector material.

\subsection{\label{sec:SuperCDMS} SuperCDMS and CDMSlite}

The Super Cryogenic Dark Matter Search experiment (SuperCDMS) operated an array of 15 interleaved Z-sen\-si\-tive ionization and phonon (iZIP) Ge detectors \cite{:RefA} from 2012 to 2015 in the Soudan Underground Laboratory to search for NRs from dark matter interactions \cite{:CDMSlite1,:HT}. Each detector was equipped with four phonon and two charge readout channels on each of the flat faces. One channel of each type acted as an outer guard ring on each side to reduce background by identifying and removing events at high radius. When operated in their normal iZIP mode with a modest bias voltage of a few volts, applied between charge and phonon sensors, simultaneous readout of phonon and charge signals enabled an effective ER-background identification for recoil energies larger than $\sim$8\,keV~\cite{:RefD}. This provided world-leading sensitivity among all solid-state detectors to WIMP masses $>$\,12\,GeV/c$^2$ \cite{:HT}.

Sensitivity to interactions of low-mass dark matter particles ($<$6\,GeV/c$^2$) was enhanced by operating one of the detectors in an alternative mode. In the CDMS low ionization threshold experiment (CDMSlite), a larger bias voltage of $\sim$70\,V was applied between the two flat faces of the detector. In this mode, the detector no longer has the capability to discriminate ER events from NR events. However, the Neganov-Trofimov-Luke mechanism \cite{:Neg,:Luke} amplifies the charge signal (in proportion to the voltage bias) into a large phonon signal, without a corresponding increase in electronic noise. In this way a much larger signal-to-noise ratio is achieved, lowering the threshold to well below a keV and thus gaining sensitivity to dark matter particles with masses of a few GeV/c$^2$. Further details on searches for low-mass dark matter with CDMSlite can be found in Refs.~\cite{:CDMSlite1, :CDMSlite2, :CDMSlite2Long}. The next-generation experiment SuperCDMS SNOLAB will further extend the low-mass experimental reach by operating new detectors (Si and Ge) based on the CDMSlite concept but optimized to achieve even lower energy thresholds ({\it HV detectors})~\cite{:sensitivity,:HVeV}.

\subsection{\label{sec:Cosmogenics} Cosmogenic Background in CDMSlite}

For CDMSlite and SuperCDMS SNOLAB, ERs from cosmogenic isotopes produced in the detector crystals during detector fabrication, testing, and storage above ground are a significant source of background. A cosmogenic isotope is of concern if its half-life is long enough that it does not decay away between the time the detectors are brought underground and the start of the dark matter search, but short enough that the decay rate is comparable to other sources of background. Half-lives of isotopes relevant to our analysis range from $\sim$100 days to a few tens of years. Table~\ref{tab:isotopes} lists all isotopes with half-lives in the relevant range that could potentially be produced in germanium by cosmogenic radiation. In addition, we include $^{71}$Ge and $^{68}$Ga. The latter has a very short half-life but is produced by the decay of the long-lived $^{68}$Ge, while $^{71}$Ge is produced during calibration measurements with a $^{252}$Cf neutron source through neutron capture on $^{70}$Ge \cite{:CDMSlite2}. 

A number of publications (listed in Table~\ref{tab:pubprod}) discuss cosmogenic activation in germanium. For a review of cosmogenic production rates in various materials, including germanium, see Ref.~\cite{:cebrianreview}. As Table~\ref{tab:pubprod} demonstrates, the different published calculations are not always in agreement with one another or with the sparse experimental results. 

Tritium ($^3$H) produced by cosmogenic radiation in germanium is expected to be the dominant background for the SuperCDMS SNOLAB HV germanium detectors~\cite{:sensitivity}. For this isotope, only one experimental result is available~\cite{:edelweiss} and the theoretical calculations show a relatively large spread in predicted activation rates. We perform a calculation in Section~\ref{sec:Calculations}, addressing some of the known shortcomings of previous approaches. In Section~\ref{sec:Data} we analyze the spectrum acquired during the second run of CDMSlite~\cite{:CDMSlite2} and extract the tritium production rate in germanium in Section~\ref{sec:production}. In Section \ref{sec:discussion}, we evaluate rates for several other isotopes either identified in CDMSlite data or reported by other experimental efforts.

\section{\label{sec:Calculations}{Cosmogenic Activation}}

\begin{table*}[htb]
\begin{center}
\small
\caption{\label{tab:isotopes}Isotopes with half-lives between 100 days and 15 years that could potentially be produced through cosmogenic activation in germanium. $^{68}$Ga and $^{71}$Ge are also included, as $^{68}$Ga is a daughter product of $^{68}$Ge and $^{71}$Ge is produced in-situ during $^{252}$Cf neutron calibrations. Half-lives are given in years (y), days (d) or minutes (m). Decay types and their branching ratios (BR) are given, including decays via electron capture (EC) directly to the ground state (GS), EC to excited states (ES), and decays via $\beta^+$ or $\beta^-$ emission. The most common gamma rays ($\gamma$) that accompany EC decays to ES, and their BRs, are also listed (511~keV gamma rays from positron annihilation are produced in pairs, thus branching ratios $>$100\,\% are possible). Q-values are also given. Isotope data are taken from Ref.~\cite{:isotopes}.}
\makebox[\linewidth][c]{%
{\renewcommand{\arraystretch}{1.5}
\begin{tabular}{|c|c|c|c|c|c|c|c|}
\hline
\multirow{2}{*}{Isotope} & \multirow{2}{*}{Half life} &\multicolumn{4}{|c|}{Decay Type(s) + BR [\%]} & $\gamma$-radiation [keV] & Q-value \\
\cline{3-6}
        &           & EC (GS) & EC (ES) & $\beta^+$ & $\beta^-$ & (branching ratio) & [keV]\\
 \hline
 \hline
$^{71}$Ge & 11.4 d  & 100 & & & & & 232.6\\
\hline
$^{68}$Ge & 270.3 d & 100 & & & & & 107.2\\
\hline
$^{68}$Ga & 68 m & 8.9 & 2.2 & 88.9 & & 511 (176\,\%), 800 (0.4\,\%), & 2921\\[-0.4ex]
          &      &     &     &        & & 1078 (3.5\,\%) & \\
\hline
$^{65}$Zn & 244.3 d & 49 & 49 & 1.7 & & 1116 (51\,\%) & 1352\\
\hline
$^{60}$Co & 5.3 y & & & &100 & 1173 (99.85\,\%), & 2823 \\[-0.4ex]
          &       & & & &    & 1333 (99.98\,\%)  & \\
\hline
$^{57}$Co & 271.9 d & & 100 & & & 14 (9.54\,\%), 122 (85.6\,\%), & 836.3\\ [-0.4ex]
          &         & &     & & & 136 (10.6\,\%), 692 (0.02\,\%) & \\
\hline
$^{55}$Fe & 2.73 y & 100 & & & & & 231.1 \\
\hline
$^{54}$Mn & 312 d & & 100 & & & 835 (100\,\%) & 1377 \\
\hline
$^{49}$V & 330 d & 100 & & & & & 601.9\\
\hline
$^{44}$Ti & 51.9 y & & 100 & & & 67.9 (93.0\,\%), 78.3 (96.4\,\%), &  267.4\\[-0.4ex]
          &        & &       & & & 146.2 (0.092\,\%) & \\
\hline
$^{45}$Ca & 162 d & & & & 100 & & 259.7 \\
\hline
$^{22}$Na & 2.6 y & & 10 & 90 & & 511 (180\,\%), 1275 (100\,\%) & 2843 \\
\hline
$^{3}$H & 12.32 y & & & & 100 & & 18.59 \\
\hline
\end{tabular}}}
\end{center}

\end{table*}

The energy transferred by cosmic radiation to an atomic nucleus may cause protons, neutrons, or nuclear clusters to escape from the core nuclear potential, dispersing the absorbed energy, and producing radioactive isotopes such as those listed in Table~\ref{tab:isotopes}.  In principle, the production rate of isotopes, $R$, by cosmic ray secondaries (neutrons, protons, muons, and pions), dominated by the contribution from neutrons, can be calculated from the production cross section excitation functions, $\sigma$, and measured cosmic-ray flux spectra, $\Phi$, for cosmic ray energy $E$ as
\begin{equation}
R = \sum_{i=n,p,\mu,\pi} \int \sigma_i \Phi_i dE_i.
\end{equation}
In practice, values for the isotope-production excitation functions rely heavily on extrapolations using nuclear models since measurements are often unavailable.  The previous efforts at calculations listed in Table~\ref{tab:pubprod} vary depending on the particular nuclear models used.  This section reevaluates these models and recalculates expected production rates for tritium and other isotopes observed in CDMSlite, as well as tritium from neutron spallation in silicon. 

\subsection{Excitation Functions from Neutron Spallation}
\label{ExcitationFunctions}

In order to understand the effect of cosmogenic radiation, we need nuclear models that describe how energy is transferred within a nucleus, how particles are ejected from an excited nucleus, and what residual nucleus remains once the energy is dissipated.  For nuclear excitation energies below $\sim$100~MeV, most particle emission occurs relatively slowly and the excitation energy is able to equilibrate among the internal degrees of freedom of the nucleus.  At higher energies, some nucleons escape before the nucleus reaches thermal equilibrium, and nucleons need to be modeled individually. In Ref.~\cite{:cebrian}, this difference in excitation function behavior at low and high energies was recognized, and appropriate codes were benchmarked and used in each region to estimate the production of mid-mass radioisotopes.

However, as tritium is produced as an ejectile rather than as a residual nucleus, models that account for clustering of ejected nucleons are required, and the tools used in Ref.~\cite{:cebrian} cannot be applied. At excitation energies below 100~MeV, detailed models for thermalized decay mechanisms and approximations to pre-equilibrium behaviour are required. TALYS is one of several codes available to implement these models and is widely used \cite{talys}, including for the activation calculations in references \cite{:mei} and \cite{:amare}. To more accurately model processes for spallation at energies of hundreds of MeV, the Liège Intranuclear Cascade model (INCL) implements Monte Carlo algorithms to simulate energy cascading amongst nucleons, and to predict how escaping nucleons cluster into nuclear fragments \cite{:INCL}. INCL comes packaged with the ABLAtion code \cite{:ABLA07}, which performs calculations similar to TALYS once the nucleus is thermalized.  Ref.~\cite{:david} compares available experimental data to a wide range of available spallation models, and INCL4.5-ABLA is shown to be significantly better than other models at predicting the production of residual nuclei from the spallation of iron, a mid-mass nucleus analogous to the germanium and silicon targets considered here.

\begin{table*}[hbt]
\begin{center}
\small
\caption{\label{tab:pubprod}Published calculated (calc.) and experimental (exp.) cosmogenic production rates for $^3$H, $^{55}$Fe, $^{65}$Zn and $^{68}$Ge in Ge. The first and second calculated values from Ref.~\cite{:avignone} use cosmic spectra from Lal \cite{:Lal} and Hess \cite{:Hess}, respectively. Ref.~\cite{:cebrian} uses cosmic spectra from Ziegler \cite{:ziegler} and Gordon \cite{Gordon}. The different calculations from Ref.~\cite{:edelweiss} use cross sections from a semi-empirical model \cite{:Tsao1,:Tsao2,:Tsao3,:Tsao4,:Tsao5} and from the MENDL-2P database \cite{:MENDL}. Calculations from Ref.~\cite{:wei} use (a) Geant4 and (b) Activia 1/2. The experimental limit for $^{68}$Ge reported in Ref.~\cite{:edelweiss} is extracted assuming full saturation of $^{68}$Ge at ground level at the time the crystal was grown. A lower $^{68}$Ge concentration at this time would imply a higher production rate. Calculations from Table 1 of Ref. \cite{:ma} use CRY to estimate the cosmic ray flux for Beijing at sea level and G\textsc{eant}4 to determine the resulting activation.}
{\renewcommand{\arraystretch}{1.5}
\begin{tabular}{|lr|c|c|c|c|c|}
\cline{4-7}
\multicolumn{3}{c}{} & \multicolumn{4}{|c|}{Production Rate [atoms/(kg$\cdot$day)]}\\[-0.4ex]
\hline
\multicolumn{2}{|c|}{Reference} &\hspace*{-0.5em}Method\hspace*{-0.5em} & $^{3}$H & $^{55}$Fe & $^{65}$Zn & $^{68}$Ge \\[-0.4ex]
 \hline
 \hline
Avignone (1992) & \cite{:avignone} & calc. & 178, 210 & - & 24.6, 34.4 & 22.9, 29.6 \\[-0.6ex]
\multicolumn{2}{|c|}{} & exp. & - & - & 38 $\pm$ 6 & 30 $\pm$ 7\\[-0.4ex]
\hline
Klapdor (2002) & \cite{:klapdor} & calc. & - & 8.4 & 79 & 58.4 \\[-0.4ex]
\hline
Barabanov (2006) & \cite{:barabanov} & calc. & - & - & - & 80.7 \\[-0.4ex]
\hline
Back (2007) & \cite{:back} & calc. & - & 3.4 & 29.0 & 45.8 \\[-0.4ex]
\hline
Mei (2009) & \cite{:mei} & calc. & 27.7$^{\dagger}$ & 8.6 & 37.1 & 41.3 \\[-0.4ex]
\hline
Cebrian (2010) & \cite{:cebrian} & calc. & - & 8.0, 6.0 & 77, 63 & 89, 60 \\[-0.4ex]
\hline
Zhang (2016) & \cite{:zhang} & calc. & 48.32 & - & - & - \\[-0.4ex]
\hline
EDELWEISS (2017)\hspace*{-1em} & \cite{:edelweiss} & calc. & 46, 43.5  & 3.5, 4.0 & 38.7, 65.8 & 23.1, 45.0 \\[-0.6ex]
\multicolumn{2}{|c|}{} & exp. &  82 $\pm$ 21 & 4.6 $\pm$ 0.7 & 106 $\pm$ 13 & >71 \\[-0.4ex]
\hline
Wei (2017) & \cite{:wei} & calc. (a)& 47.37 & 4.47 & 24.93 & 21.75 \\[-0.4ex] 
 &  & calc. (b)& 33.70/51.27 & 1.13/1.18 & 5.34/9.66 & 7.04/15.38 \\[-0.4ex] 
\hline
Amare (2018) & \cite{:amare} & calc. & 75 $\pm$ 26 & - & - & - \\[-0.4ex] 
\hline
Ma (2018) & \cite{:ma} & calc. & 23.68 & 4.15 & 40.47 & 83.05 \\[-0.4ex] 
\hline
\multicolumn{7}{l}{\footnotesize$^{\dagger}$See text (section~\ref{ExcitationFunctions}) and footnote~\ref{fn_mei} for a discussion of this value.}
\end{tabular}}
\vspace{0.5cm}
\end{center}
\end{table*}%

For the estimates presented here, we use a slightly newer version of this code (INCL++-ABLA version 5.2.9.5) with its default parameters, and TALYS version 1.8 with custom parameters. \footnote{Some parameters whose default values help reduce computation time were relaxed.  Specifically, \texttt{maxlevelstar}, \texttt{maxlevelsres}, and \texttt{maxlevelsbin} for all light ejectiles up to mass 4 were increased to 40 to account for known nuclear levels that may affect nucleon production, the pre-equilibrium model contribution was calculated for all incident energies, and thresholds for discarding negligible reaction channels, \texttt{xseps} and \texttt{popeps}, were reduced to $10^{-15}$.  For all other parameters, default values were used.} Cross sections calculated with TALYS are used for neutron energies below 100~MeV, and those calculated with INCL++-ABLA are used above neutron energies of 100~MeV.

\begin{figure}[htb]
\centering
\includegraphics[angle=270]{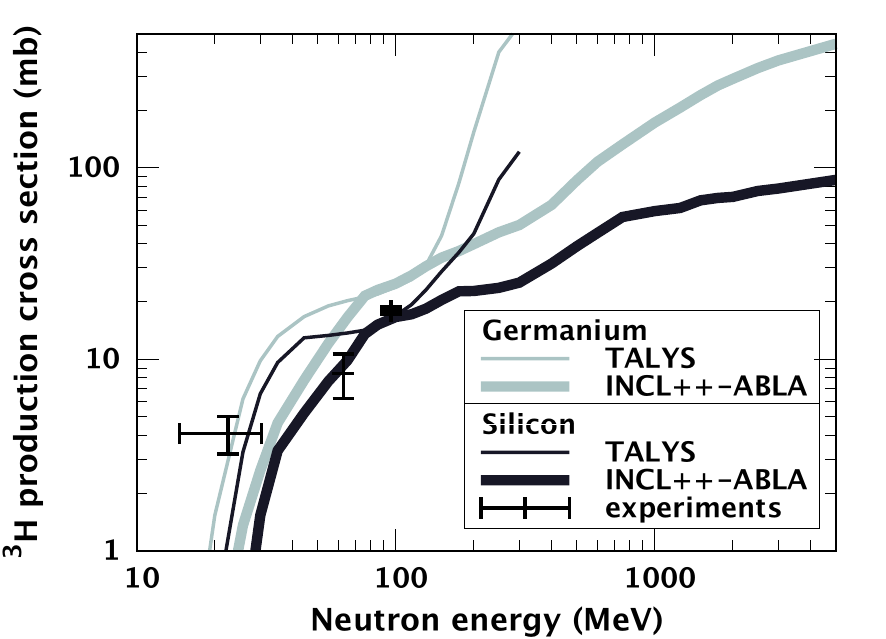}
\caption{\label{fig:prod_xs} Tritium production cross sections for neutron spallation on $^{\text{nat}}$Ge and $^{\text{nat}}$Si targets calculated using TALYS version 1.8 \cite{talys} and INCL++-ABLA version 5.2.9.5 \cite{:INCL, :ABLA07}.  Results from experiments that measured the tritium production cross section in $^{\text{nat}}$Si are also shown \cite{PhysRevC.69.064609, QAIM1978150, Benck}.  The later calculations in this section use the TALYS cross sections below 100 MeV and the INCL++-ABLA cross sections above 100 MeV.}
\end{figure}

\begin{figure}[htb]
\centering
\includegraphics[angle=270]{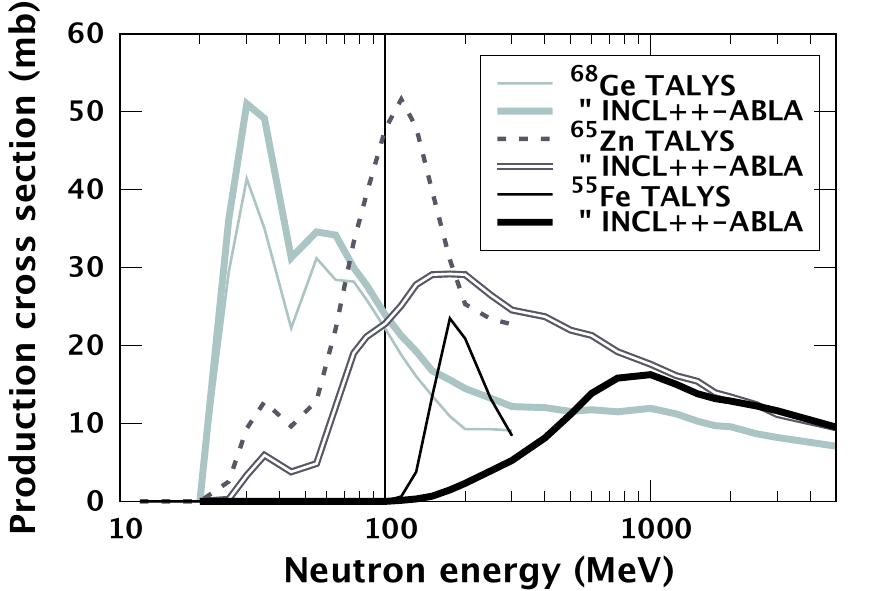}
\caption{\label{fig:ge_xs} Calculated production cross sections for neutron spallation on $^{\text{nat}}$Ge using TALYS 1.8 \cite{talys} and INCL++-ABLA version 5.2.9.5 \cite{:INCL}. TALYS cross sections are used below 100 MeV (vertical line) and the INCL++-ABLA cross sections above 100 MeV.}
\end{figure}

Figure~\ref{fig:prod_xs} shows the calculated tritium production excitation functions in Ge and Si with natural isotopic composition ($^{\text{nat}}$Ge and $^{\text{nat}}$Si, respectively). The same method was used to produce, in Figure~\ref{fig:ge_xs}, the production excitation functions of the other isotopes listed in Table~\ref{tab:pubprod}.  For these isotopes, the excitation functions have similar shapes to those in Ref.~\cite{:cebrian} using complementary methods, but are generally slightly lower.

As a check, the general shape of the isotope production excitation functions can be inferred before running the calculations.  For most isotopes, the cross section peaks near a small multiple of the nucleon separation energy ($\sim$10~MeV per ejected nucleon) then falls as the number of alternative exit channels in the reaction increases.  For tritium, which may be emitted multiple times during nuclear deexcitation, the production cross section grows monotonically and sub-linearly with the collision energy, from threshold to energies on the order of the total nuclear binding energy ($\sim1$~GeV). Note that in Ref.~\cite{:mei} TALYS was used to calculate a tritium production cross section that did not increase monotonically, thus significantly reducing the calculated production rate. \footnote{\label{fn_mei}An attempt to calculate the tritium production cross sections using TALYS 1.0 as used in Ref.~\cite{:mei} did not reproduce their result for neutron energies above 80~MeV. In addition, in Ref. \cite{:mei} the exposure of the IGEX detector crystals is overstated by nearly a factor of nine \cite{:cebrian_priv}, leading to an apparent, but \emph{false}, confirmation of their calculated value.}

Studies to benchmark the TALYS and INCL-ABLA models have demonstrated accuracies of better than 40\,\% in most of their respective domains of applicability for reactions similar to those considered above.  TALYS has been benchmarked to the measured production of residual nuclei from proton irradiation at various energies, as well as from neutron irradiation up to 180~MeV in Si, Co, Fe, Ni, and Cu targets \cite{:michel}.  The latter study shows disagreements for the production of light residual nuclei that require multiple emissions in their production, such as a factor of $\sim$5 overprediction for $^{48}$Cr production from an iron target.  By restricting TALYS to excitation energies below 100~MeV, this multiple emission regime is partially avoided.  Benchmark studies of INCL-ABLA calculations for the production of residual nuclei with $Z$ between 13 and 24 by proton irradiation at 300~MeV of an $^{56}$Fe target show that predictions are generally within 40\,\% of the measured values \cite{:david}.  INCL-ABLA fails to accurately predict the production of the isotopes $^{54}$Co and $^{52,53,54}$Fe, suggesting that it may not be suitable for calculating the cross section of processes that do not require charged particle emission from the target, such as the production of $^{68}$Ge from spallation in germanium.  Fortunately, the cosmogenic production of such isotopes is dominated by neutrons with energies below 100~MeV for which TALYS provides reasonably accurate excitation functions.  By using TALYS for excitation energies below 100~MeV and INCL-ABLA at higher energies, the known failures of these models are avoided, and an uncertainty of $\pm40\,\%$ can be propagated to the predicted cosmogenic production rates.

As the excitation functions of both TALYS and INCL-ABLA agree at 100~MeV for most of the cosmogenic radioisotopes considered in Figures~\ref{fig:prod_xs} and \ref{fig:ge_xs}, the exact choice of 100~MeV versus other nearby energies contributes negligibly to uncertainties in the predicted cosmogenic production rates.  However a significant difference is observed at a neutron energy of 100~MeV in the production of  $^{65}$Zn.  Other choices for this cutoff between 20~MeV and 300~MeV may change the predicted production rate by up to $\pm20\,\%$, still small compared to the considered $\pm40\,\%$ uncertainty.

Calculations of tritium production from neutron spallation can be compared to measurements using 96~MeV neutrons on silicon \cite{PhysRevC.69.064609} and iron \cite{PhysRevC.70.014607}.  The calculations and experiments agree within the small 5\,\%-level experimental uncertainties for both TALYS and INCL-ABLA calculations, with calculated production cross sections in iron of 21.8 and 21.6 mb respectively versus a measurement of 21$\pm$1.1 mb.  Despite the good agreement for these specific data points, the overall uncertainty on the calculations for tritium should not be considered more precise than the typical uncertainty of $\sim$40\,\% observed in general.

\begin{figure}[htb]
\centering
\includegraphics[angle=270]{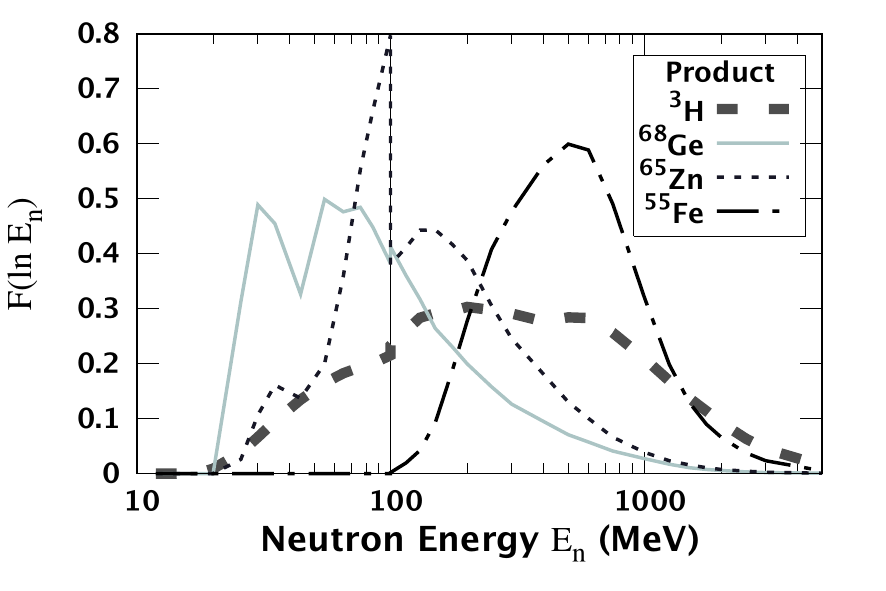}
\caption{\label{fig:ge_importance} Calculated probability distribution functions in log-energy for the neutron energy associated with cosmogenic isotope production in $^{\text{nat}}$Ge. These distribution functions, $F(\ln E_n)$, 
are given by the product of the cross sections shown in Figure \ref{fig:ge_xs} and the energy-dependent cosmogenic neutron flux. The vertical line indicates the switch from TALYS to INCL-ABLA cross sections.
} 
\end{figure}

\subsection{Predicted Cosmogenic Activation Rates}

Several competing parameterizations of the sea-level neutron flux exist, as noted in Ref.~\cite{:cebrianreview}. We adopt the model of Gordon~\cite{Gordon} for consistency with other recent estimates \cite{:mei,:amare, :cebrian}. Figure~\ref{fig:ge_importance} uses the excitation functions of Figures~\ref{fig:prod_xs} and \ref{fig:ge_xs} and the adopted sea level neutron spectrum to show the expected contribution of neutrons of different energies to the production of particular radioisotopes. 

The cosmic-ray neutron fluxes published in Ref.~\cite{Gordon} are normalized to the average cosmic-ray flux observed at sea level in New York. Adjustments for solar cycle variation \footnote{The solar cycle correction was obtained using data from the neutron monitor at Newark/Swarthmore.  These are provided by the University of Delaware Department of Physics and Astronomy and the Bartol Research Institute.  These data were accessed using the NMDB database at www.nmdb.eu.}, altitude, and the latitude-dependent geomagnetic cutoff were considered, but it was found that for the location and time period of above-ground fabrication and storage of the CDMSlite detector --- Stanford University, from 2009 to 2011 --- these percent-level corrections largely cancel, so the New York sea-level normalization from Ref. \cite{Gordon} is used.

Forms of radiation other than fast neutrons may also cause transmutation into the isotopes listed in Table~\ref{tab:isotopes}.  The most important of these is spallation by cosmic-ray protons. In the energy range that contributes most to the production of cosmogenic radioisotopes, from 0.1 to 1~GeV, the proton flux is $\sim$5\,\% of the neutron flux. At these energies, the spallation processes induced by protons and neutrons are very similar; thus, the calculated production cross sections from neutrons have been increased by 5\,\% to approximately account for the proton flux.  Other publications find contributions of 3\% to 25\% \cite{:barabanov,:zhang,:wei,:amare,:ma} for the isotopes considered herein.

In addition to spallation processes, cosmogenic activation can occur from stopped negative muon and pion capture.  Approximately 500~muons/(kg$\cdot$day) are stopped in materials at the Earth's surface, and at shallow depths up to 5 meters of water equivalent \cite{:charalambus}.  The capture of these negative muons converts a proton into a neutron while releasing tens of MeV into the nucleus. Ref.~\cite{:wyttenbach} reports the measured fraction of these captures that generate various residual isotopes.  This provides a small ($\mathcal{O}$(1\,\%)) addition to the production rate of tritium and some other radioisotopes at the earth's surface with production energy thresholds below 100~MeV. We ignore this contribution in our calculated rates; however, it may be important for the production of cosmogenic radioisotopes for materials stored for long periods in sites with shallow overburden, where the production rate from cosmic-ray neutrons is substantially reduced.  This process was also considered in \cite{:barabanov} for the production of $^{60}$Co from germanium and found to be negligible.  

The total calculated production rates in $^{\text{nat}}$Ge are 95 atoms/(kg$\cdot$d) for $^3$H, 5.6~atoms/(kg$\cdot$d) for $^{55}$Fe, 51~atoms /(kg$\cdot$d) for $^{65}$Zn, and 49~atoms/(kg$\cdot$d) for $^{68}$Ge; these values are also listed in Table \ref{tab:prodrate}. The calculated production rate of $^3$H in $^{\text{nat}}$Si is 124~atoms/(kg$\cdot$d).

\section{\label{sec:Data} Experimental Analysis of CDMSlite Run 2}

In this section, we reanalyze the CDMSlite Run 2 spectrum (originally used in Ref.~\cite{:CDMSlite2} to search for low-mass WIMPs) using a likelihood method to extract background event rates due to cosmogenically produced radioisotopes. A background model is constructed that includes the tritium beta-decay spectrum, a relatively flat component due to scattering of higher energy gamma rays (mostly from radioactive contaminants such as the U, Th chains and $^{40}$K in the experimental setup) with incomplete energy transfer (``Compton background''), and several peaks.  The latter are produced by X-ray/Auger-electron cascades following electron-capture (EC) decays of radio-isotopes to the ground states of their daughter nuclei. Table \ref{tab:EC} lists the total cascade energies and branching ratios (BR) for captures from different shells for the EC-decay isotopes that we consider. We include all those listed in Table \ref{tab:isotopes} except for $^{22}$Na and $^{44}$Ti, for which there is no evidence in the CDMSlite spectrum. Potential contributions from non-tritium beta decays with higher-energy endpoints are not explicitly considered, but are accounted for in the fit by the Compton background contribution (see Section \ref{section:systematics}). The known above-ground exposure history of the detector is then used to convert statistically significant detections from the likelihood fit to cosmogenic production rates.

The prior analysis of the CDMSlite Run 2 spectrum included energies only up to 2~keV \cite{:CDMSlite2}, including evaluation of the detection efficiency. All of the EC decays that we consider dominantly give rise to peak energies above 2~keV (cf.\ K-shell captures in Table~\ref{tab:EC}). Furthermore, to effectively differentiate between spectral contributions from tritium betas and the Compton background, the likelihood fit should include energies above the tritium beta-decay endpoint. Consequently, an important aspect of the analysis presented here is an extension of the published CDMSlite detection efficiency to higher energies.

\begin{table*}[hbt]
\small
\begin{center}
\caption{\label{tab:EC}Data for the radioisotopes considered in this work that decay via electron capture: total X-ray/Auger-electron cascade energies $E$ (keV), measured experimental resolutions $\sigma$ (eV) at those energies \cite{:CDMSlite2Long}, and branching ratios (BR) \cite{:peaks} for the K-, L$_1$-, and M$_1$-shell capture peaks.
For isotopes other than $^{71}$Ge and $^{68}$Ge only these three peaks are relevant due to the low number of decays of these isotopes and the small branching ratios to other shells. However, for germanium the L$_2$ peak ($E=1.14$~keV, $\sigma=29$~eV, and BR~=~0.1\,\%) cannot be neglected due to the high rate of Ge EC decays.}
{\renewcommand{\arraystretch}{1.5}
\begin{tabular}{|c|ccc|ccc|ccc|}
\hline
\multirow{2}{*}{Isotope}  & \multicolumn{3}{|c|}{K} & \multicolumn{3}{|c|}{L$_1$} & \multicolumn{3}{|c|}{M$_1$}   \\
\cline{2-10}
&E [keV] &$\sigma$ [eV] & BR &E [keV] &$\sigma$ [eV] & BR &E [keV] &$\sigma$ [eV] & BR \\
 \hline
 \hline
 $^{71}$Ge & 10.37 & 101 & 87.6\,\% & 1.30 & 31.2 & 10.5\,\% & 0.160 & 14.0 & 1.8\,\% \\
 $^{68}$Ge & 10.37 & 101 & 86.5\,\% & 1.30 & 31.2 & 11.5\,\% & 0.160 & 14.0 & 1.78\,\% \\
 $^{68}$Ga & 9.66 & 96.3 & 88.6\,\% & 1.20 & 30.0 &  9.8\,\% & 0.140 & 13.1 & 1.6\,\% \\
 $^{65}$Zn & 8.98 & 91.8 & 88.6\,\% & 1.10 & 28.8 &  9.8\,\% & 0.122 & 13.1 & 1.6\,\% \\
 $^{57}$Co & 7.11 & 79.2 & 88.8\,\% & 0.84 & 25.5 &  9.6\,\% & 0.091 & 12.3 & 1.5\,\% \\ 
 $^{55}$Fe & 6.54 & 75.2 & 88.6\,\% & 0.77 & 24.4 &  9.8\,\% & 0.082 & 12.1 & 1.6\,\% \\
 $^{54}$Mn & 5.99 & 71.3 & 89.6\,\% & 0.70 & 23.4 &  9.0\,\% & 0.066 & 11.7 & 1.4\,\% \\
 $^{49}$V  & 4.97 & 63.7 & 89.3\,\% & 0.56 & 21.3 &  9.3\,\% & 0.059 & 11.5 & 1.4\,\% \\
\hline
\end{tabular}}
\end{center}
\end{table*}%

\subsection{\label{section:efficiency}CDMSlite Detection Efficiency above 2~keV}

\begin{figure}[htb]
  \centering       
   \includegraphics[width=0.5\textwidth, keepaspectratio]{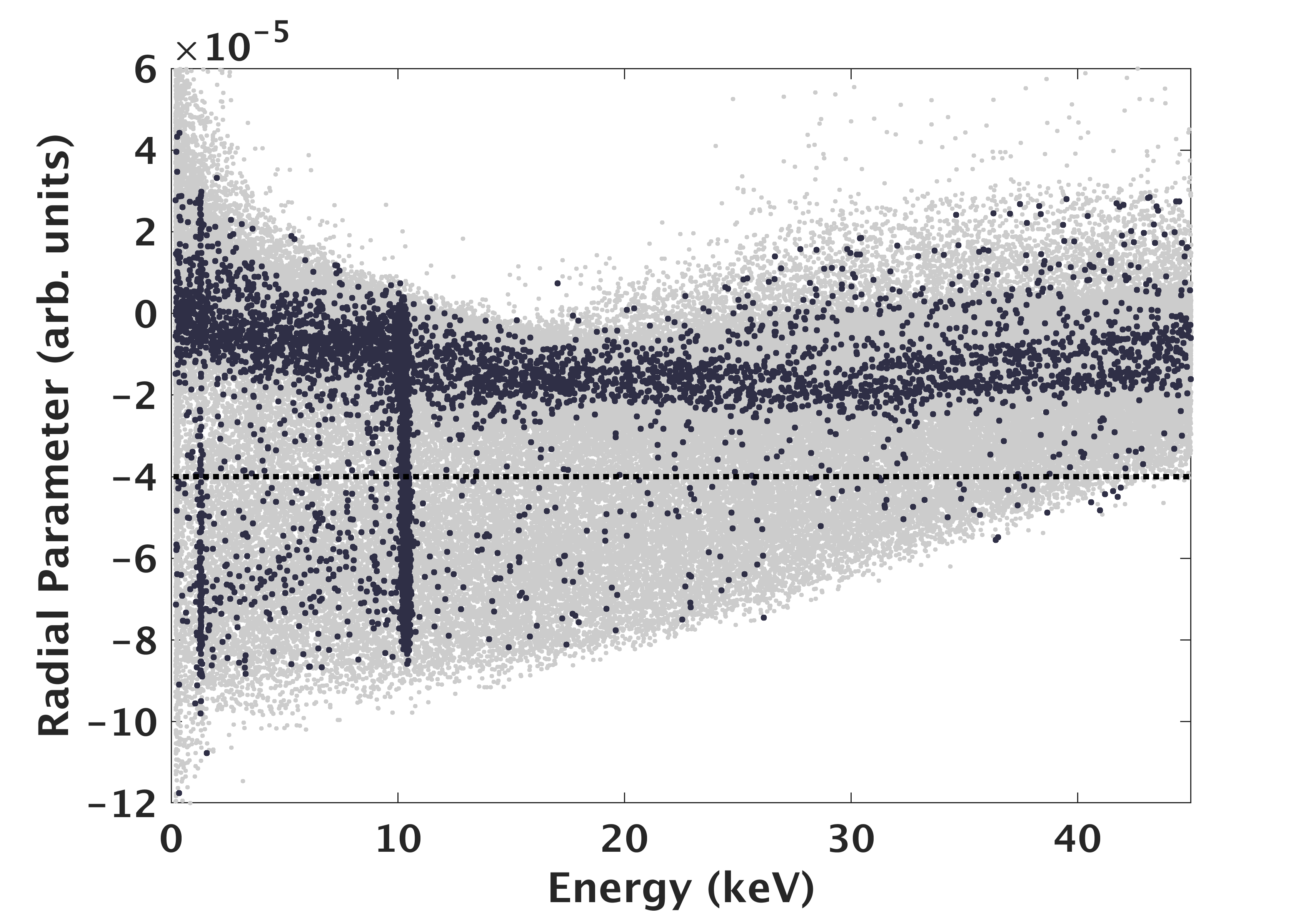}
             \caption{\label{fig:radvsenergy}Radial parameter (for details see Ref. \cite{:CDMSlite2Long}) vs.\ energy for part of CDMSlite Run 2, showing dark matter search data (dark points) overlaid on $^{133}$Ba $\gamma$-calibration data (light grey). Events falling below the radial cut (dotted line) are considered to be within the fiducial volume and are selected for analysis. It is evident that the selection efficiency decreases with increasing energy above $\sim$20~keV, which is particularly obvious for the $^{133}$Ba event distribution.}
   \end{figure}

For CDMSlite Run 2 the efficiency above 100~eV is of order 50\,\% and is largely determined by the radial fiducialization ({\it radial cut}) that is necessary to remove data from regions of the detector where an inhomogeneous electric fields leads to a reduced Neganov-Trofimov-Luke amplification and thus a significantly distorted energy spectrum \cite{:CDMSlite2,:CDMSlite2Long}. Using the $^{71}$Ge capture lines and a simulation method based on pulse shape \cite{:ryan}, the radial cut efficiency was shown to be fairly flat below the 1.3~keV L-shell line. For energies directly above the 1.3 keV line, up to about 2~keV, there is no indication that the radial event distribution changes significantly; the radial cut efficiency is thus linearly interpolated from 1.3~keV to 10.37~keV for energies between 1.3~keV and 2~keV. However, at energies above about 2~keV the outer phonon channel shows partial signal saturation, leading to a reduction in efficiency of the radial fiducialization \cite{:CDMSlite2Long}. This downward trend is confirmed by an estimate of the efficiency using events from the 10.37~keV K-shell line \cite{:CDMSlite2Long}. The decreasing selection efficiency with increasing energy can be observed in Figure \ref{fig:radvsenergy}, which shows the distribution of the radial parameter (on which the radial cut is based) as a function of energy.

Based on the observed distribution, we start our study with an initial hypothesis for the detection efficiency over the full energy range of interest (threshold to 20~keV) defined as follows: 

\begin{itemize}
\item Below 2~keV the previously published efficiency is used.

\item From 2 to 10.37~keV we assume that the efficiency drops linearly down to 45.4\,\% --- the efficiency reported in Ref.~\cite{:CDMSlite2Long} for the Ge K-shell line.

\item Above 10.37~keV the efficiency is presumed to be constant. This is a simple choice based on the behaviour of the radial distribution below $\sim$20~keV, and is not expected to account for the decreasing selection efficiency at higher energies.

\end{itemize}

\begin{figure}[tb]
  	\centering
          \includegraphics[width=0.5\textwidth, keepaspectratio]{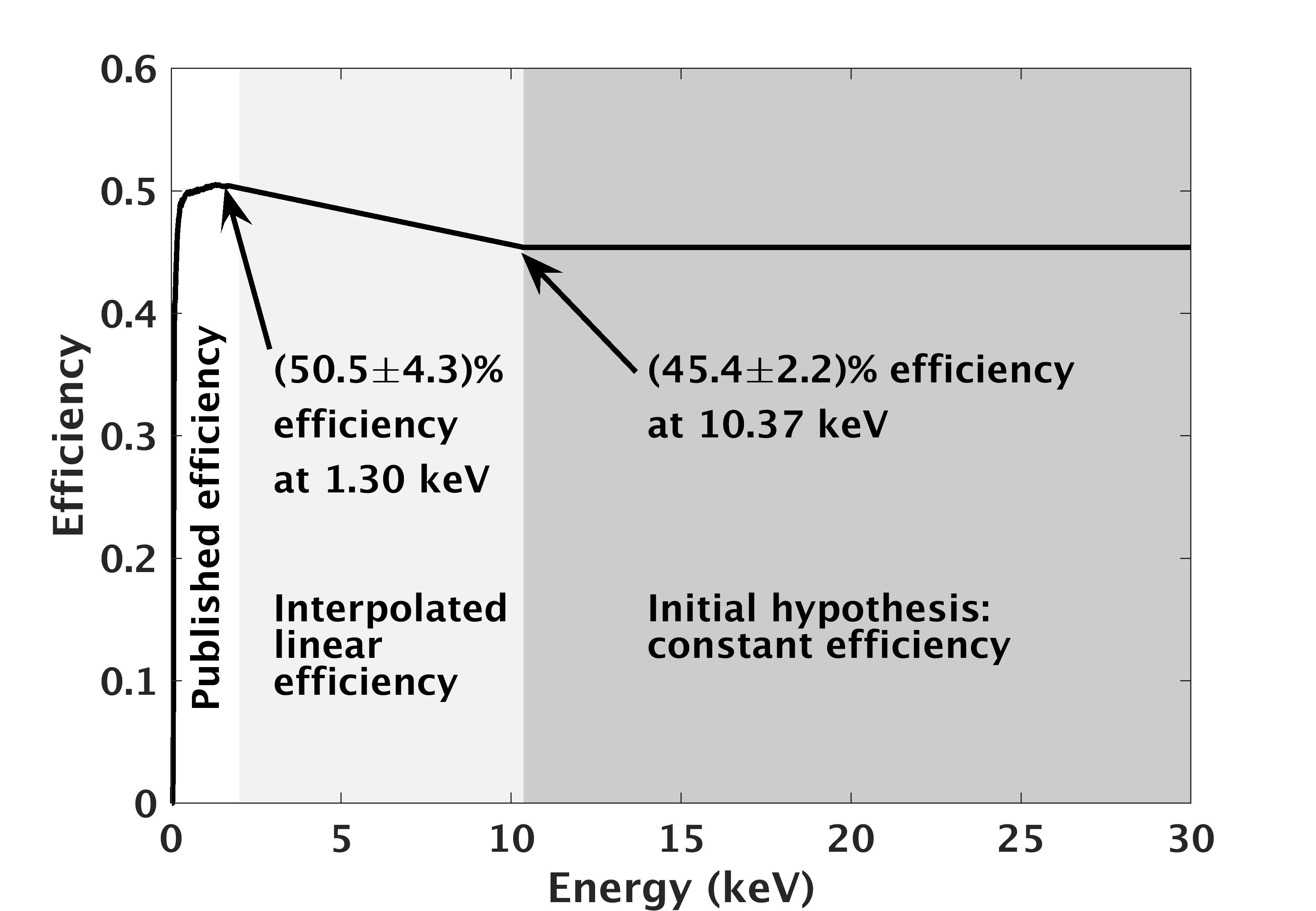}   
               \caption{\label{fig:2kevEff} Initial hypothesis for the CDMSlite detection efficiency as a function of energy (black line). Below 2~keV it matches the published efficiency curve \cite{:CDMSlite2Long} (white region). We linearly interpolate from 2~keV to the value reported in Ref.~\cite{:CDMSlite2Long} at 10.37~keV (light grey region). Above 10.37~keV, we use a constant efficiency as our initial estimate (dark grey region).}
\end{figure}

Figure \ref{fig:2kevEff} shows this initial estimate of the efficiency function.

\begin{figure}[htb]
  	\centering
    \includegraphics[width=0.5\textwidth, keepaspectratio]{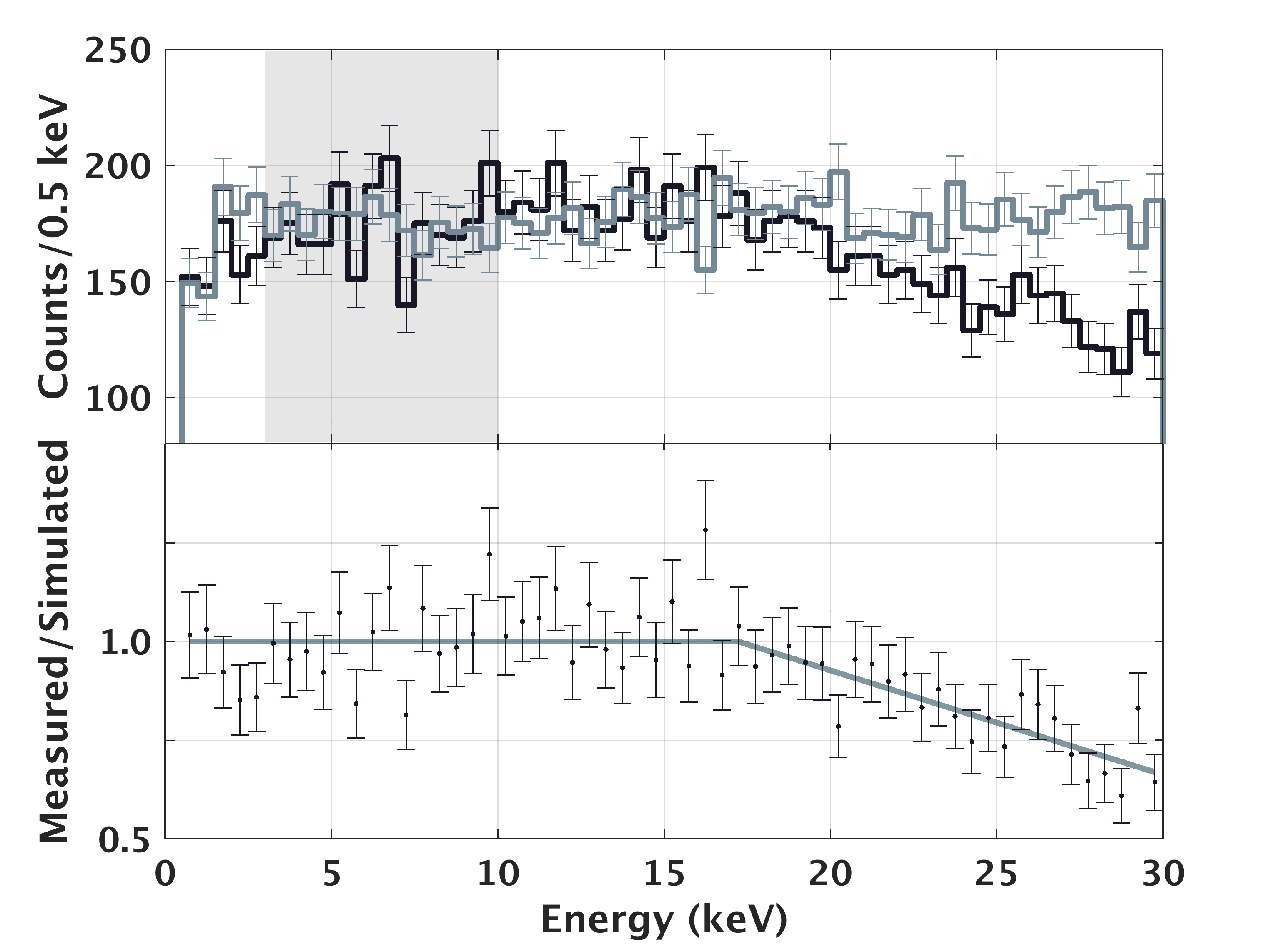}  
               \caption{\label{fig:effcorrfit}{\it Top}: Comparison of the measured (dark histogram) and simulated (light histogram) $^{133}$Ba calibration spectra, where the initial efficiency hypothesis shown in Figure \ref{fig:2kevEff} has been applied to the latter. The gray shaded region corresponds to the energy range from 3 to 10~keV, which was used to normalize the simulated spectrum to the measured rate.  Error bars correspond to 1$\sigma$--uncertainties.\\
{\it Bottom}: Ratio of the measured to simulated energy spectra from the top panel (points), compared to the best-fit, piecewise defined efficiency correction function $f(E)$ (solid line).}
\end{figure}

In order to test this initial hypothesis we compare $^{133}$Ba $\gamma$-calibration data to a Monte Carlo simulation generated for the same experimental configuration using G\textsc{eant}4 \cite{:geant4,:geant4_2,:geant4_3,:geant4update}. The initial-hypothesis efficiency is applied to the simulated energy spectrum, which is then normalized to the corresponding measured rate in the energy range between 3 and 10~keV. The top panel of Figure~\ref{fig:effcorrfit} shows the resulting simulation together with the measured spectrum. The two spectra are in good agreement below $\sim$18~keV, thus supporting the initial efficiency hypothesis. However, there is a significant discrepancy above $\sim$20~keV, growing with increasing energy, that reflects the diminishing performance of the radial parameter in this energy range (as seen in Figure~\ref{fig:radvsenergy}).

We derive a correction to the initial efficiency hypothesis based on the ratio of the measured to simulated spectra. As shown in the bottom panel of Figure ~\ref{fig:effcorrfit}, this ratio decreases approximately linearly with increasing energy in the upper portion of the energy range. Therefore, we introduce as a correction a piecewise defined function of energy $f(E)$, which is constant (unity) below some energy $E_0$ and decreases linearly with a slope $S$ above this energy. The values of the parameters $E_{0}$ and $S$ are determined by fitting this correction function to the ratio of measured and simulated spectra in the energy range from 0.5 to 30~keV, as shown in Figure~\ref{fig:effcorrfit} (bottom panel).  The best-fit values and their 95\,\% C.L. uncertainties are  (17.3\,$\pm$\,2.7) keV and ($-$0.026\,$\pm$\,0.009) keV$^{-1}$, respectively. A reduced $\chi^2$ of 0.84 indicates that this is a good fit.

\begin{figure}[htb]
  	\centering
          \includegraphics[width=0.5\textwidth, keepaspectratio]{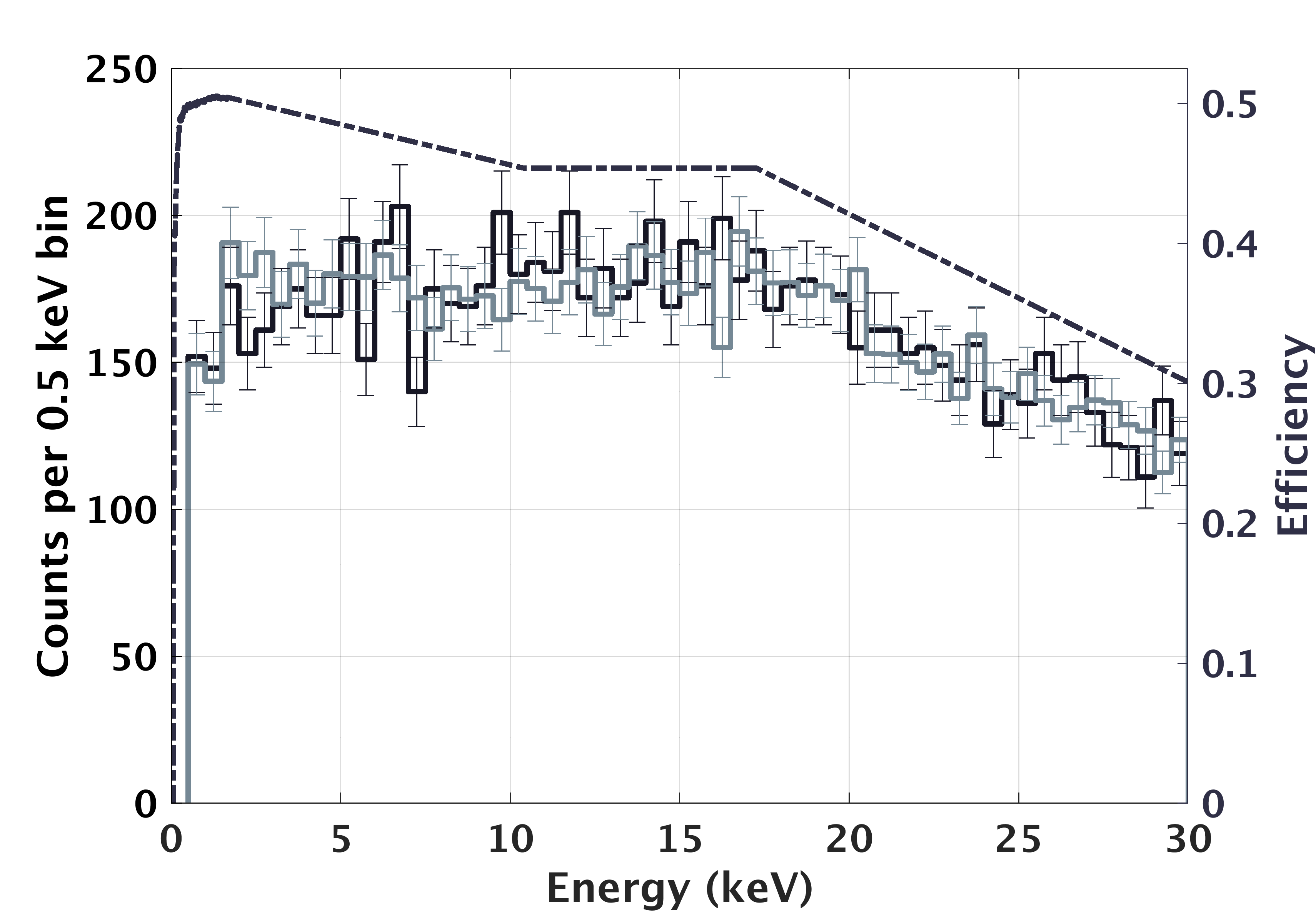}   
               \caption{\label{fig:finaleff} Comparison of $^{133}$Ba calibration data (dark histogram) with the G\textsc{eant}4 simulation, corrected by the final efficiency and normalized to the data in the range from 3 to 10~keV (light histogram). Also shown is the final efficiency function (dashed line, right axis).}
\end{figure}

Figure~\ref{fig:finaleff} shows the final, corrected detection efficiency over the full energy region considered.  Also shown are the measured and simulated $^{133}$Ba calibration spectra, where the final efficiency has been applied to the latter. Use of a Kolmogorov-Smirnov test \cite{:ks} to compare the measured and simulated spectra, both before and after application of the correction function $f(E)$, confirms that the corrected efficiency is a more accurate representation of the true detection efficiency. With p values of 1.6$\times 10^{-4}$ and 0.78 for the initial and corrected efficiencies, respectively, the initial hypothesis is rejected  at 99.9\,\% confidence, while the final efficiency is consistent with the measured spectrum.

The uncertainty on the final efficiency is determined by propagating the uncertainties on the initial efficiency (determined analogously to the efficiency itself) with the fit uncertainties of $f(E)$, leading to a maximum relative uncertainty of $\sim$8\,\%.

To gauge how the choice of the 3~keV lower bound of the energy normalization range impacts the results, this value was varied between 0.5 and 4.5~keV, resulting in a maximum variation of the efficiency of $\sim$3.5\,\% (relative). This is sub-dominant compared to the uncertainty for the final efficiency function discussed above.

\subsection{Analysis of the CDMSlite Spectrum}

In order to determine the contributions of the different components to the CDMSlite Run 2 spectrum, a maximum likelihood fit is performed. The likelihood analysis includes models for EC X-ray peaks, the tritium beta-decay spectrum, and a component corresponding to interactions of higher energy gamma rays from the decay of radio-contaminants in the setup (e.g. U, K, and Th) depositing only a fraction of their energy (hereafter referred to as the ``Compton'' component). 

The energy spectrum of each component --- EC peaks, tritium and Compton --- is modeled by a probability distribution function (PDF), to which the final efficiency determined in Sec. \ref{section:efficiency} is applied), with an associated likelihood estimator corresponding to the number of events that the component contributes to the overall spectrum. For $N$ events, with energies denoted by $E_i$, the negative log-likelihood function is

\begin{equation}
-\ln(\mathcal{L}) = \sum_b n_b - \sum_{i=1}^N \ln(\sum_{b} n_bf_b(E_i))
\end{equation}

\noindent where $f_b(E_i)$ are the individual background PDFs and $n_b$ are the number of events that each background contributes to the spectrum.

\subsubsection{Electron-Capture Peaks}

The EC peaks of each radioisotope in Table \ref{tab:EC} are modeled by Gaussian functions centered at the K-, L$_1$- and M$_1$-shell binding energies of the respective daughter isotope, with the standard deviation of the Gaussian set by the energy-dependent resolution function reported in Ref.~\cite{:CDMSlite2Long} and listed in Table \ref{tab:EC}. In the likelihood fit, the amplitude ratios between the K-, L-, and M-shell peaks for each radioisotope are fixed according to the expected branching ratios in Ref.~\cite{:peaks} (listed in Table \ref{tab:EC}), ignoring potential uncertainties. 

The L$_2$-shell contribution is neglected for all isotopes other than germanium, as the branching ratio is on the order of $\sim$0.1\,\% (compared to the L$_1$-shell branching ratio on the order of $\sim$10\,\%). As $^{71}$Ge makes a significant contribution to the spectrum, the germanium L$_2$-shell is included in this analysis.    

\subsubsection{Tritium-Beta Decay Spectrum}

The tritium beta-decay spectrum is given by 
\begin{equation}
N(T_e) = C\sqrt{T_e^2 + 2T_em_ec^2}(Q-T_e)^2(T_e+m_ec^2)F(Z,T_e),
\end{equation}
\noindent where $C$ is a normalization constant, $T_e$ is the kinetic energy of the emitted electron (i.e.\ the energy measured by our detector), $m_e$ is the mass of the electron, and $Q$ is the Q-value \cite{:krane}. For the Fermi function, $F(Z,T_e)$, where $Z$ is the atomic number of the daughter nucleus, we use the following non-relativistic approximation \cite{:povh}:
\begin{equation}
F(Z,T_e) = \frac{2\pi \eta}{1-e^{-2\pi \eta}},
\end{equation}
\noindent where $\eta = \alpha Z\frac{c}{v}$ with the fine structure constant $\alpha$, and $v$ is the electron velocity. This spectrum is convolved with the energy-dependent resolution function.

\subsubsection{Compton Background Component}

The spectral shape of the Compton model is simulated with G\textsc{eant}4 based on the Monash model \cite{:Monash,:dannCompton}. The Monash model takes into account changes to the gamma-ray scattering rate that occur at small scattering angles where the energy transfer is of order of the atomic binding energies. Steps at the germanium K-, L-, and M-shell binding energies appear as fewer and fewer electrons are available for the scattering process, as shown in Figure~\ref{fig:ML_spec}.  

\subsubsection{Likelihood Fit Results} 
\label{sec:fit}

\begin{figure}[htb]
  \centering
          \includegraphics[width=0.5\textwidth, keepaspectratio]{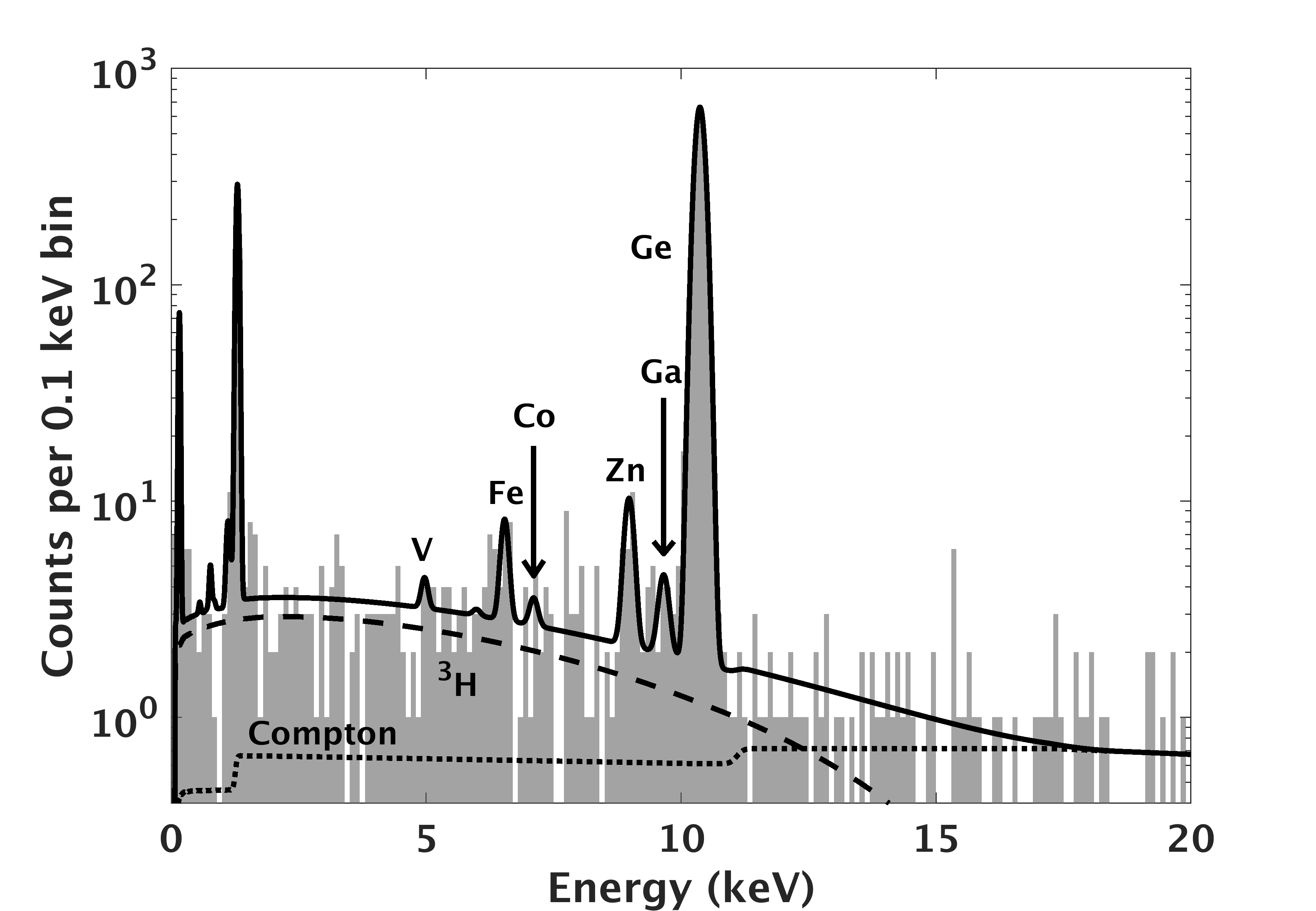}
     \caption{\label{fig:ML_spec}Maximum likelihood fit to the CDMSlite Run 2 spectrum. The L- and M-shell peaks are not labeled, but occur in the same order as the K-shell peaks.}
   \end{figure}

\begin{figure}[htb]
  \centering
\includegraphics[width=0.5\textwidth, keepaspectratio]{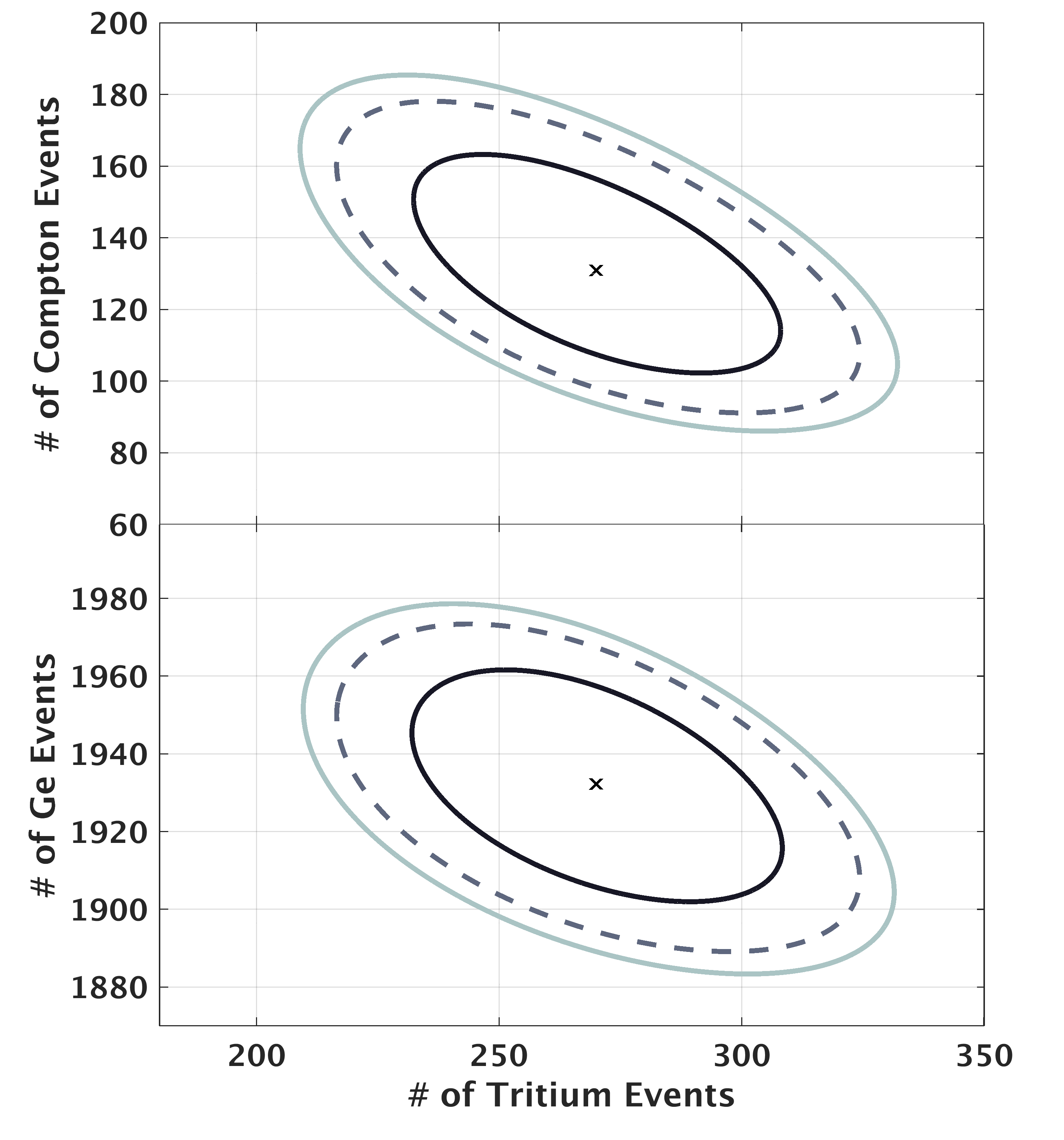}
     \caption{\label{fig:corr}Correlation between the event counts of tritium and Ge (top), and those of tritium and Compton (bottom). Contours are drawn at the 95$^{\mathrm{th}}$ (outer, light), 90$^{\mathrm{th}}$ (middle, dashed), and 68$^{\mathrm{th}}$ (inner, dark) percentiles. The best-fit values are indicated ($\times$).}
   \end{figure}

The results of the likelihood fit are shown in Figure\ \ref{fig:ML_spec}. The uncertainty on each fit parameter is determined from its likelihood distribution by varying the value of the parameter over a wide range about the best-fit value, calculating the likelihood at each value. The uncertainties are then extracted from the resulting likelihood distribution. Similarly, we also calculated the 2-dimensional correlations. The two examples with the strongest correlation (tritium vs.\ Compton and tritium vs.\ Ge) are shown in Figure\ \ref{fig:corr}, indicating that the uncertainties on the other components only have a small effect on the tritium result. The fit results are summarized in Table \ref{tab:ml}. All values refer to the number of events contributed by the respective component to the measured spectrum.  

Other Ge-based rare event searches have identified additional isotopes such as $^{49}$V, $^{54}$Mn, $^{56}$Co, $^{57}$Co, $^{58}$Co, $^{60}$Co, $^{63}$Ni, and $^{67}$Ga \cite{:avignone, :klapdor, :barabanov, :mei, :cebrian, :zhang, :edelweiss, :wei, :amare}. Thus, the fit includes not only those isotopes for which there is clear evidence in the CDMSlite data, but also three additional isotopes: $^{49}$V, $^{54}$Mn, and $^{57}$Co. All three have half lives within the relevant range. The fit values for these isotopes (also included in Table \ref{tab:ml}) are compatible with zero. The remaining isotopes are neglected: $^{56}$Co and $^{58}$Co have half-lives that are too short and, in addition, would not be distinguishable from $^{57}$Co; the same argument applies for $^{67}$Ga (indistinguishable from $^{68}$Ga and having too short of a half-life). $^{60}$Co and $^{63}$Ni are $\beta^-$ emitters ($^{63}$Ni with a half-life of 101.2~y, therefore not included in Table \ref{tab:isotopes}) with Q-values well above our energy region of interest, thus contributing an almost flat component that absorbed in the Compton component. Section \ref{section:systematics} further motivates neglecting these beta emitters. As mentioned in Section \ref{sec:Cosmogenics}, $^{22}$Na and $^{44}$Ti have appropriate half-lives and could potentially be produced cosmogenically in germanium, but inclusion of these isotopes in the fit results in negligible amplitudes for both. 

\begin{table*}[hbt]
\begin{center}
\small
\caption{\label{tab:ml} Number of events that each component contributes to the measured CDMSlite spectrum with 70.10 kg-days of exposure \cite{:CDMSlite2}, as determined by the maximum likelihood fit. The lower limit (LL) and the upper limit (UL) are given for each confidence level (CL). For the isotopes in the last three rows, there is no evidence for their presence, which is clear from the negative lower bounds on their confidence intervals.}

{\renewcommand{\arraystretch}{1.5}
\begin{tabular}{|c|c|cc|cc|cc|}
\hline
 &  & \multicolumn{6}{|c|}{Uncertainty Range} \\
\cline{3-8}
Component  & \# Events & \multicolumn{2}{c|}{$95\,\%$ CL} & \multicolumn{2}{c|}{$90\,\%$ CL} & \multicolumn{2}{c|}{$68\,\%$ CL}\\[-1ex]
 & & LL & UL & LL & UL & LL & UL \\
 \hline
 \hline
$^{68,71}$Ge & 1932 & 1893 & 1967 & 1899 & 1962 & 1912 & 1949 \\
$^{68}$Ga & 7.2 & 0.9 & 18.0 & 1.8 & 16.1 & 3.9 & 12.6 \\
$^{65}$Zn  & 21.5 & 11.9 & 35.1 & 13.4 & 32.8 & 16.6 & 28.3 \\
$^{55}$Fe  & 11.5 & 3.8 & 23.6 & 5.0 & 21.6 & 7.7 & 17.7 \\
$^{3}$H  & 270 & 222 & 318 & 230 & 310 & 245 & 294 \\
Compton  & 131 & 95 & 175 & 101 & 168 & 113 & 153 \\  
\hline
\hline
$^{58,57,56}$Co & 2.0 & $-$2.7 & 11.2 & $-$2.0 & 9.6 & $-$0.3 & 6.7 \\
$^{54}$Mn & 0.4 & $-$3.7 & 9.2 & $-$3.1 & 7.7 & $-$1.7 & 4.9 \\
$^{49}$V & 2.2 & $-$2.2 & 10.7 & $-$1.5 & 9.2 & 0.2 & 6.5\\
\hline
\hline
Sum  & 2378 & & & & & & \\
\hline
\end{tabular}}
\end{center}
\end{table*}%

\subsubsection{Time-Dependence of the EC Rates}

The half-lives of the observed EC decays from cosmogenic isotopes are between 240~days and about 3~years (see Table \ref{tab:isotopes}). As a result, over the course of the measurement period of roughly one year, the measured rate in the EC peaks is expected to drop. We have studied the time dependence, and in all cases the rate as function of time is consistent with the respective decay time, but due to the small number of observed events, the half-lives cannot be positively confirmed. No constraint on the decay of the EC peaks was used in the likelihood analysis of Section \ref{sec:fit}. 

The only case where a clear time dependence is observed is the decay of the Ge EC peak, which is dominated by the decay of $^{71}$Ge produced \textit{in-situ} by neutron activation during three nuclear recoil calibration campaigns separated by several months \cite{:CDMSlite2}. In principle, the time dependence of the rate in the Ge EC peak could provide an additional way to extract the $^{68}$Ge decay rate. However, the time distribution and strength of the $^{71}$Ge signal together with the overall measurement schedule, with a significant gap for maintenance of the cryogenic equipment in the summer of 2014, led to a large uncertainty in this analysis. The constraints on the $^{68}$Ge decay from the time dependence are considerably weaker than (but compatible with) those deduced using likelihood fit results for the $^{68}$Ga EC peak.

\subsection{\label{section:systematics}Systematic Uncertainties from the Choice of the Background Model}

Before drawing conclusions from the fit results about the cosmogenic production rates of the observed isotopes, it is important to understand how the presence of unidentified background components could impact those fit results.

\subsubsection*{\texorpdfstring{$\beta^-$}{} Decays}
In addition to tritium there are other $\beta$-active nuclei that can be produced cosmogenically. However, all of the isotopes that can be produced and fall within the relevant decay-time window have considerably higher endpoint energies. This means that the contribution in the energy range of interest is reduced accordingly and that their spectra are close to flat. Therefore our fit would absorb them in the Compton contribution. Literature values for the production rate of $^{60}$Co and $^{63}$Ni are available and range from 2.0 to 6.6 and from 1.9 to 5.2 atoms/(kg$\cdot$day) respectively \cite{:mei,:klapdor}. Even assuming the highest values, they would make up only a few percent of the deduced Compton contribution and thus can be safely neglected as separate terms in the likelihood fit.

\subsubsection*{$\beta^+$ Decays}

For both $^{68}$Ga and $^{65}$Zn, $\beta^+$-decay is an alternative to the previously considered EC, with branching ratios of 88.9\,\% and 1.7\,\%, respectively (see Table \ref{tab:isotopes}). However, the expected combined contribution of these $\beta^+$ backgrounds to the measured spectrum below 20~keV is less than one event.

\subsubsection*{Instrumental Noise}

The instrumental noise background is effectively removed by the analysis \cite{:CDMSlite2Long} and is therefore ignored here.

\subsubsection*{Surface events from $^{210}$Pb} 

Surface events may be a non-negligible component of the observed spectrum. The spectral shape of this background depends critically on the geometrical distribution of this contaminant but is generally expected to rise in the energy range below a few keV~\cite{:CDMSII_LTLikelihood}. If such a component is present in the data (but ignored in the fit) it would lead to an over-estimate of the tritium rate. A detailed study of a potential contribution of this background to the data discussed here has not yet been carried out, but estimates based on the observed alpha rates in the detector suggest that it contributes not more than about 8\,\% to the continuous low-energy spectrum. A correction to the extracted tritium rate would be subdominant compared to the statistical uncertainty.

\subsubsection*{Unidentified backgrounds}

There is no indication of significant contributions to the observed spectrum from other sources. However, since the cosmogenic tritium is expected to be the dominant background in the Ge detectors of SuperCDMS SNOLAB~\cite{:sensitivity}, it is important to understand how unidentified background could impact the conclusion about the tritium production rate. The two most extreme assumptions about unidentified background in this context would be a background that has a shape similar to the tritium spectrum or a background that dominates the spectrum at high energy but drops to zero in the range where the tritium may contribute. In the former case we could explain the spectrum without the presence of tritium while the latter case provides a very conservative upper limit for the tritium rate and thus can be used for a conservative prediction of the expected sensitivity of SuperCDMS SNOLAB.

\begin{figure}[H]
  \centering
          \includegraphics[width=0.5\textwidth, keepaspectratio]{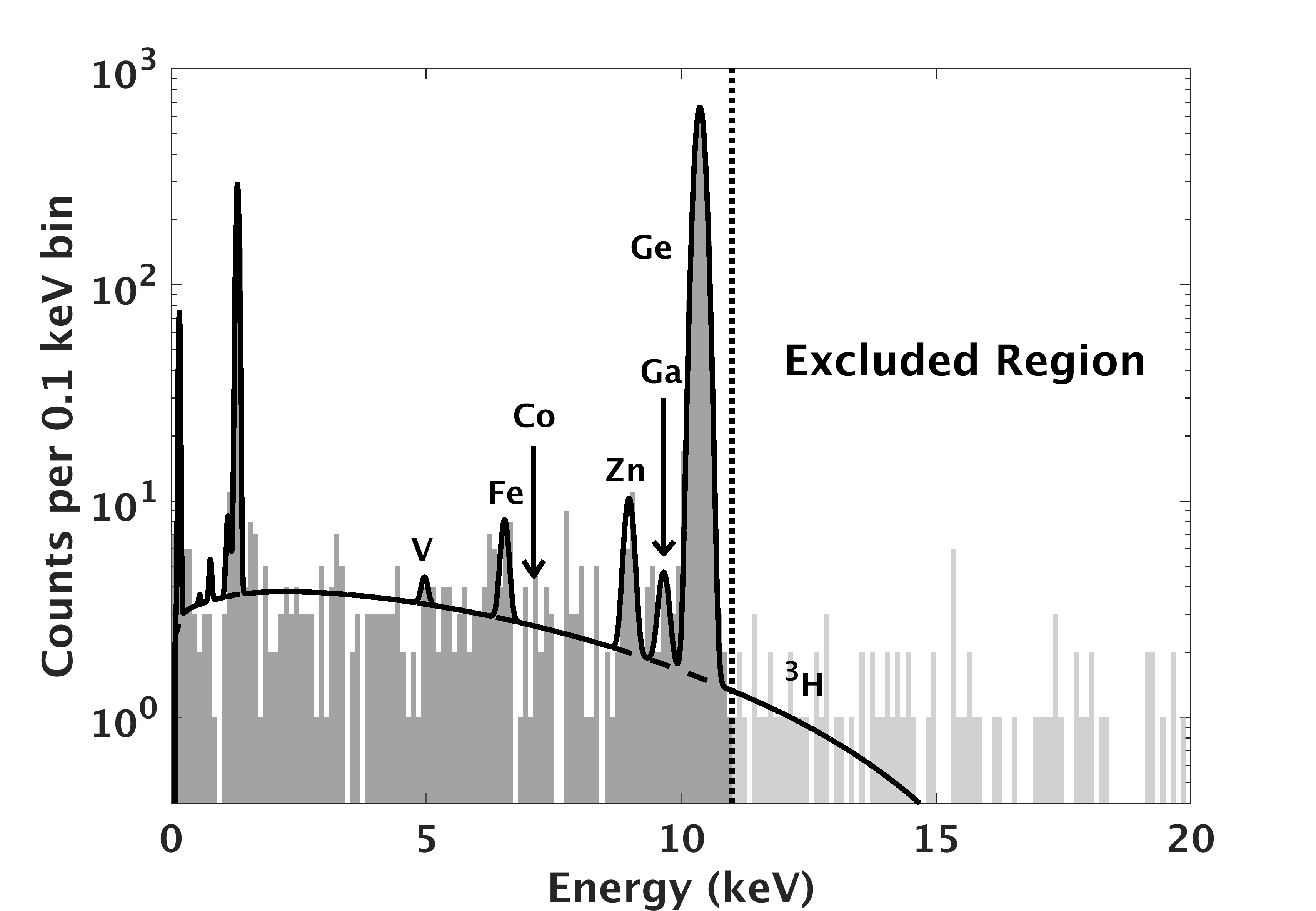}
     \caption{\label{fig:ML_noComp}Maximum likelihood fit neglecting Compton background to determine a conservative upper limit on $^3$H. The fit region is limited to below 11~keV (vertical dotted line) in this case, since the $^3$H spectrum alone cannot reproduce the data at higher energy.}
   \end{figure}

In order to produce such a conservative estimate, we performed a second likelihood fit where the Compton component is set to zero. Because a pure tritium spectrum is incompatible with the observed spectral shape near the endpoint, this fit is performed over a restricted energy range, only up to 11~keV. The result of this fit is shown in Figure\ \ref{fig:ML_noComp}. The extracted tritium rate in this case is roughly 30\,\% higher than for the best fit discussed earlier.

\section{\label{sec:production}Experimental Production Rates}

\subsection{Efficiency Correction for Gamma Emitting Isotopes}
\label{sec:GammaCorr}

Both $^{65}$Zn and $^{68}$Ga can decay via EC to an excited state of the daughter nucleus, releasing a $\gamma$ in the subsequent transition to the ground state. These decays only appear in the EC peaks if the $\gamma$ escapes the {CDMSlite} detector without interaction; if the gamma-ray {\it does} interact in the CDMSlite detector, it will shift the event's energy out of the EC peak. If the gamma-ray escapes the CDMSlite detector but strikes another operating detector in the same tower, the event is classified as a multiple-scatter event and removed as part of the standard dark matter analysis event selection. In both of these cases the number of events in the EC peak is reduced when compared to the decay rate of the respective isotope. The spectrum discussed above (Figures \ref{fig:ML_spec}, \ref{fig:ML_noComp}) only includes single-scatter events as they are derived from the SuperCDMS standard WIMP event selection criterion \footnote{One may expect the EC peaks of these two isotopes to also appear in the multiple-scatter spectrum. However, we confirmed using G{\sc eant}4 Monte Carlo simulations that the probability for the gamma to leave the CDMSlite detector without interaction and subsequently interact in a neighbouring detector is too small for a resulting feature to be visible in our multiple-scatter data.} 
As it is our goal to determine cosmogenic production rates for the various isotopes, we will consider these inefficiencies in more detail. We use data from a G\textsc{eant}4 simulation to determine the fraction of events removed from the measured EC peaks due to a $\gamma$ interaction in the same or another detector. 

The simulation model is the same as used in \cite{:sensitivity}, but adapted for the experiment at the Soudan Underground Laboratory and modified to simulate the decays of $^{65}$Zn and $^{68}$Ga in the CDMSlite detector. An analysis of the simulation output, mirroring that of the CDMSlite dark matter analysis, shows that (64.3\,$\pm$\,0.1)\,\% of $^{65}$Zn events and (9.64\,$\pm$\,0.04)\,\% of $^{68}$Ge events are expected to appear in their respective single-scatter EC peak. The given uncertainties are from simulation statistics. 

\subsection{Detector History}

The CDMSlite detector has a well documented location history. After crystal pulling on November 24$^{\mathrm{th}}$, 2008 at ORTEC, in Oak Ridge, TN, the detector spent 1065~days above ground during the fabrication and testing process at various locations in the San Francisco Bay Area (including Berkeley, Stanford, and SLAC), with intermittent storage periods in a shallow underground tunnel at Stanford that has shielding of 16~m water equivalent \cite{:tunnelC} against cosmic rays. Subsequently the detector was brought to the Soudan Underground Laboratory (1100~m water equivalent) on October 25$^{\mathrm{th}}$, 2011. CDMSlite Run 2 started 833~days later on February 4$^{\mathrm{th}}$, 2014, and took place over a period of 279~days.   

If a detector is exposed for a time much longer than the life-time of a given isotope, it will eventually reach saturation with a constant decay rate for atoms of this species determined by the cosmogenic production rate. 
Given the measurement schedule, detector history and the life-time of the respective isotope we appropriately integrate the production and decay equations to convert the measured number of decays, using the detection efficiency, into a production rate in atoms per kg of detector material per day of exposure to comic radiation. The production rate is assumed to be constant during the times the detector was exposed to cosmic radiation. Corrections are made for the EC decays accompanied by $\gamma$ emission, as discussed in Section \ref{sec:GammaCorr}. For tritium we additionally determine a conservative upper limit using the result from the second likelihood fit that neglects the Compton background and thus attributes all events between the EC peaks to the tritium spectrum. 

\subsection{Production Rates}
It is assumed that all cosmogenic isotopes, with the exception of $^{68}$Ge, are expelled during the pulling of the crystal. For $^{68}$Ge we make two extreme assumptions: either the amount of this isotope at the time of pulling is zero, or the crystal is already in full saturation. We calculate the production rate for both of these extreme assumptions. Since the exposure history after pulling is long compared to the life-time of the isotope and the rate of observed $^{68}$Ga events from which the $^{68}$Ge activity is deduced is rather small, the effect of this uncertainty in the final production rate is small compared to the statistical uncertainty. Both values are listed in Table \ref{tab:prodrate} together with the results for all other isotopes. The calculated production rates from Section \ref{sec:Calculations} are also listed for comparison.

While the detector history during detector production and testing is well documented, there is some uncertainty in the travel history of the detector which is important for elevation and shielding. As a conservative approach, the maximum uncertainties in the recorded detector history are considered and then propagated with the uncertainties (68$^{\text{th}}$ percentile) from the likelihood fit (cf.\ Table \ref{tab:ml}), detection efficiency (cf.\ Section \ref{section:efficiency}), and efficiency for gamma emitting isotopes (cf.\ Section \ref{sec:GammaCorr}). 

\begin{table}[H]
\begin{center}
\small
\caption{\label{tab:prodrate}Production rates and 1$\sigma$ uncertainties for different isotopes in natural germanium at sea level. The second number for $^3$H (in parenthesis) is deduced from the fit that neglects contributions from Compton scattering, giving a very conservative upper limit for the $^3$H production rate. The two values for $^{68}$Ge make the two extreme assumptions that the concentration of this isotope during crystal pulling was either zero or in equilibrium (saturated).}
{\renewcommand{\arraystretch}{1.5}
\begin{tabular}{|c|c|c|c|}
\hline
\multirow{2}{*}{Isotope} & \multicolumn{3}{|c|}{Cosmogenic Production Rate [atoms/(kg$\cdot$d)]} \\
\cline{2-4}
& Calculation & Measurement & Comment\\ 
 \hline
 $^{3}$H & 95 & 74 $\pm$ 9 & best fit\\
 & & (97 $\pm$ 10) & no Compton\\
 $^{55}$Fe & 5.6 & 1.5 $\pm$ 0.7 &\\
 $^{65}$Zn & 51 & 17 $\pm$ 5 & \\
 $^{68}$Ge & 49 & 30 $\pm$ 18 & 0 initial \\ 
 & & 27 $\pm$ 17 & saturated\\ 
 \hline
\end{tabular}}
\end{center}
\end{table}%

\section{\label{sec:discussion}Discussion and Conclusion}

With this analysis of the data from the second run of CDMSlite at Soudan we expand the knowledge base of cosmogenic production rates in natural germanium for various isotopes, including tritium. 

The best-fit tritium production rate of (74\,$\pm$\,9)~atoms /(kg$\cdot$day) determined here is slightly lower than, though within uncertainty of, the production rate of (82\,$\pm$\,21) atoms/(kg$\cdot$day) measured by EDELWEISS \cite{:edelweiss}. This holds true even if we consider potential contributions from additional backgrounds discussed in Section \ref{section:systematics} that are ignored in the main analysis, which would likely reduce the extracted tritium rate by a few \%.

The measured production rates for the other isotopes, $^{55}$Fe, $^{65}$Zn and $^{68}$Ge, however, are considerably lower (see Table \ref{tab:pubprod}) than those measured by EDELWEISS.

At first glance the CDMSlite and EDELWEISS measurements appear incompatible. However, it is conceivable that the discrepancy can be explained with a difference in the flux and spectra of the cosmogenic radiation between the two experiments and the assumption that other factors may impact the concentration of $^{68}$Ge. This is also a possible explanation for the discrepancy between the calculations and the measurements.

A conclusive interpretation of the data will likely require a better understanding of the production mechanisms, including an improved knowledge of the temporal and spatial variation of cosmogenic neutron fluxes, as well as additional well-controlled activation measurements.

\section{Acknowledgements}

The SuperCDMS collaboration gratefully acknowledges technical assistance from the staff of the Soudan Underground Laboratory and the Minnesota Department of Natural Resources. The iZIP detectors were fabricated in the Stanford Nanofabrication Facility, which is a member of the National Nanofabrication Infrastructure Network, sponsored and supported by the NSF. Funding and support were received from the National Science Foundation, the U.S.\ Department of Energy, Fermilab URA Visiting Scholar Grant No.\  15-S-33, NSERC Canada, the Canada First Research Excellence Fund, and MultiDark (Spanish MINECO).  This document was prepared by the SuperCDMS collaboration using the resources of the Fermi National Accelerator Laboratory (Fermilab), a U.S.\ Department of Energy, Office of Science, HEP User Facility. Fermilab is managed by Fermi Research Alliance, LLC (FRA), acting under Contract No.\ DE-AC02-07CH11359. Pacific Northwest National Laboratory is operated by Battelle Memorial Institute under Contract No.\ DE-AC05-76RL01830 for the U.S.\ Department of Energy. SLAC is operated under Contract No.\ DEAC02-76SF00515 with the U.S.\ Department of Energy.

\end{document}